\documentclass[%
superscriptaddress,
reprint,
showpacs,
 amsmath,
 amssymb,
 aps,
pra,
]{revtex4-1}

\usepackage{graphicx,color}
\usepackage{dcolumn}
\usepackage{bm}
\usepackage{amssymb}

\begin{document}

\title{
Cold, anisotropically-interacting van der Waals molecule: TiHe
}

\author{Nancy Quiros}
\affiliation{Department of Physics, University of Nevada, Reno, NV 89557, USA}
\author{Naima Tariq}
\affiliation{Department of Physics, University of Nevada, Reno, NV 89557, USA}
\author{Timur V. Tscherbul}
\affiliation{Department of Physics, University of Nevada, Reno, NV 89557, USA}
\author{Jacek K{\l}os}
\affiliation{Department of Chemistry and Biochemistry, University of Maryland, College Park, MD, 20742, USA}
\author{Jonathan D. Weinstein}
\email{weinstein@physics.unr.edu}
\homepage{http://www.physics.unr.edu/xap/}
\affiliation{Department of Physics, University of Nevada, Reno, NV 89557, USA}

\date{\today}


\begin{abstract}

We have used laser ablation and helium buffer-gas cooling to produce the titanium--helium van der Waals molecule at cryogenic temperatures. The molecules were detected through  laser-induced fluorescence spectroscopy. Ground-state Ti($a^3F_2$)-He binding energies 
were determined for the ground and first rotationally excited states from studying equilibrium thermodynamic properties, and found to agree well with theoretical calculations based on  newly calculated {\it ab initio}  Ti-He interaction potentials, opening up novel possibilities for studying the formation, dynamics, and non-universal chemistry of  van der Waals clusters  at low temperatures.   
\end{abstract}





\maketitle

%

Weakly bound complexes of atoms and molecules held together by long-range van der Waals (vdW) forces are key to understanding a wide range of phenomena in physics and chemistry, ranging from classical and quantum chaos \cite{Berry1992chaos} and phase transitions \cite{Mandelshtam2006phaseTransitionsClusters} to the universal physics of  Efimov trimers and quantum droplets \cite{JoseReview, Bulgac, Javier}. In condensed-phase chemical physics, vdW clusters  serve as a model to study the mechanisms of solvation, nucleation, and chemical reactivity \cite{vdwChemistry,Brahms2011VdW,tang2002quantum}. In the context of quantum many-body physics, vdW molecules can be used to explore the formation of exotic quasiparticles in superfluid helium nanodroplets \cite{LemeshkoAngulon}. Helium-containing vdW molecules are the lightest of all vdW clusters, and hence are of particular interest as model systems, in which to study the emergence of macroscopic quantum phenomena such as superfluidity  \cite{doi:10.1146/annurev.physchem.49.1.1}.  


Thus far, the experimental study of He-containing vdW molecules has focused on molecules formed in supersonic expansions  \cite{jcp.98.3564, He3Efimov, smalley1976fluorescence, stephenson1984relaxation, tang2002quantum, mckellar2006helium}.   Recent groundbreaking advances in the production and trapping of translationally cold molecules \cite{RomanNJP09} made it possible to create trapped ensembles of cold polar molecules with high enough densities to study collisions and chemical reactions \cite{RomanNJP09, JunKRb10,PhysRevLett.108.203201}  and carry out ultra-precise spectroscopic measurements to probe the physics beyond the Standard Model \cite{ThOedm}. The production and trapping of cold vdW molecules would similarly enable  highly sensitive spectroscopic detection of heretofore unobserved clusters, as well as the study and control of their quantum dynamics \cite{PhysRevLett.105.033001,Brahms2011VdW,BretislavPhysics2013,Krems04}.


We have recently observed the formation of cold, ground-state LiHe molecules in a cryogenic He buffer gas \cite{tariq2013spectroscopic}. The LiHe molecule has a single near-threshold bound state with a binding energy of 0.024 cm$^{-1}$ \cite{LiHeBindingEnergies:2002} comparable to that of the He$_2$ dimer \cite{jcp.98.3564,BretislavPhysics2013}. Due to their vanishingly small binding energies, these molecules belong to an exotic class of quantum halo dimers characterized by extremely delocalized wavefunctions, universal properties, and enormously   large three-body formation rates \cite{JensenQuantumHalos:2004,JoseReview}. As the binding energies  of most other atoms and molecules with He  are much larger than those of LiHe and He$_2$ \cite{Brahms2011VdW}, it is  far from obvious whether atom-He vdW clusters would form upon  immersing the parent atoms into cryogenic He buffer gas. In fact, no such clusterization was observed in recent buffer gas cooling experiments involving large hydrocarbon molecules \cite{JohnNonStickingStilbene}.

Here, we report on the first observation of a cold vdW molecule TiHe with a binding energy of 80 times  that of LiHe, thereby providing direct experimental evidence for the formation of deeply bound vdW molecules at  cold (rather than ultracold) temperatures of 1-2 K, where three-body recombination occurs in the much less explored non-universal and multiple partial wave regimes, opening up the prospects for the experimental study of new few-body physics   at finite collision energy \cite{WangEsryNJP2011}. 

Unlike the atom-He complexes previously studied, the TiHe molecule features anisotropic interactions due to the highly degenerate $^3F_2$ ground state of atomic Ti; these anisotropic interactions have previously only been observed in collisional experiments \cite{Hancox05TiHe,krems05TiHe,lu2008fine}.
To our knowledge, the TiHe vdW complex is the first neutral  molecular ground state described by Hund's case (e) ever detected; this state had been observed previously for molecular ions and Rydberg molecules \cite{carrington1996microwave, zhu1993spin, osterwalder2004high}. We carry out rigorous {\it ab initio} and multichannel bound-state calculations of the complex's bound levels and find quantitative agreement with experiment.



High densities of cold atomic titanium are produced by laser ablation of a solid titanium target and $^4$He buffer-gas cooling  \cite{lu2008fine} as described in the Supplemental Material \cite{TiHe2016sup}.
 Ground-state titanium atoms are detected by laser absorption spectroscopy  on the $a^3F_2\rightarrow\ y^3F_2$ transition at $25107.410$~cm$^{-1}$ \cite{NISTAtomicData}. Helium densities are measured with a pressure gauge; we have measured TiHe signal at helium  densities from  $3\times 10 ^{16}$~cm$^{-3}$ to $8\times 10 ^{17}$~cm$^{-3}$.
In this environment, we expect TiHe molecules to form by three-body recombination \cite{tariq2013spectroscopic}.
We search for the TiHe molecules produced using laser-induced fluorescence  (LIF) spectroscopy.





\begin{figure}[t]
    \begin{center}
      \includegraphics[width=\linewidth]{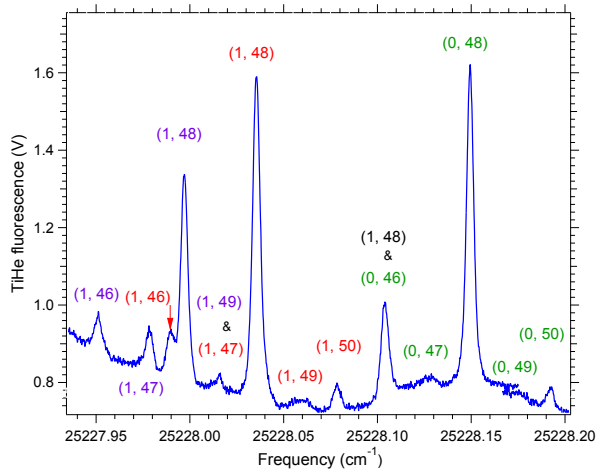}  
    \caption{ \label{fig:spectrum}
           Fluorescence spectrum of TiHe molecules taken at a helium density of $6.5\times 10 ^{17}$~cm$^{-3}$, cell temperature of $1.2$~K and a probe power of $1.20$~mW.  Not shown is an additional small peak at 25227.5~cm$^{-1}$.   The peaks are labelled as ($N$, $A$), where $N$ is the ground state rotational quantum number, and $A$ is the isotope.  Identification as described in the text.  
          The peaks at 25227.95, 25227.98, and 25228.02~cm$^{-1}$ have higher intensities than would we expected from isotopic analysis, we suspect they may be overlapped with  unidentified transitions from the $N=1$ state. The frequency center of the scan is determined from a wavemeter with an uncertainty of 0.1~cm$^{-1}$.
                                                }       
    \end{center}
\end{figure}

A spectroscopic search near the $a^3F_2\rightarrow\ y^3F_3$ atomic Ti transition at $25227.220$~cm$^{-1}$ \cite{NISTAtomicData} revealed LIF signals blue-detuned from this transition by roughly 1~cm$^{-1}$, as shown in Fig.~\ref{fig:spectrum}. All data discussed in this paper is from these peaks; additional spectral peaks are discussed in the Supplemental Material \cite{TiHe2016sup}. 

The spectrum is complicated by the presence of multiple isotopes of titanium (due to the low natural abundance of $^3$He, we expect all signals to correspond to $^4$He). Fortunately, line identification is straightforward because the observed isotope shifts closely match those of the atomic transition, as discussed in the Supplemental Material \cite{TiHe2016sup}.

Unfortunately, the excited state structure is not known sufficiently well to identify the remaining lines by their spectral patterns \cite{TiHe2016sup}. Thankfully, we are able to determine the ground state properties of the larger peaks through their equilibrium properties, and conclude that they correspond to TiHe molecules, as explained below.

We studied the dependence of the LIF signal on temperature, helium density, and titanium density ($T$, $n_\text{He}$ and $n_\text{Ti}$).
In equilibrium, the expected TiHe density for a single bound state of degeneracy $g$ is
\begin{equation}
n_\text{TiHe} = g~ n_\text{Ti} n_\text{He} \lambda_{dB}^3 \cdot e^{-E/T}
\label{eq:equil}
\end{equation}
where $E$ is the binding energy and $\lambda_{dB}$ is the reduced-mass thermal deBroglie wavelength \cite{PhysRevLett.105.033001}.



We calibrate our LIF signal with a simultaneously measured absorption signal from Ti atoms to determine the TiHe optical depth (OD), which is proportional to density. We measure the Ti OD directly via absorption spectroscopy. 

\begin{figure}[t]
    \begin{center}
      \includegraphics[width=1.12\linewidth, trim = 85 80 0 40]{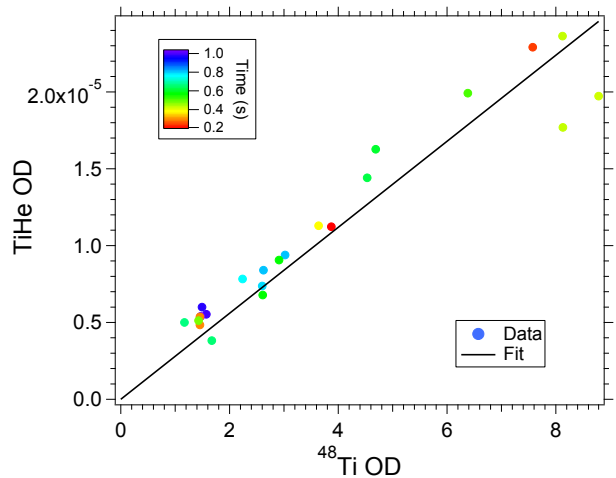}
    \caption{ \label{fig:TiHeOD_vs_TiOD_oneT} 
               TiHe OD as a function of Ti OD, colored according to the time after the ablation pulse. The TiHe OD increases linearly with Ti OD as seen by the linear fit $y=bx$.
               The data was taken at temperatures between 1.97 K to 2.10 K at $n_{He}=3.4\times 10 ^{16}$~cm$^{-3}$. 
           }       
    \end{center}
\end{figure}

We measure the temperature of the gas from the atomic Ti spectrum. The Ti linewidth has contributions from Doppler broadening, pressure broadening, and the natural linewidth.  We experimentally measure a pressure broadening coefficient of $1\times 10^{-10}$~Hz~cm$^{3}$. We deconvolve the Voigt profile of the Ti peak into its Gaussian and Lorentzian contributions \cite{olivero1977empirical} and  calculate the translational temperature for the Ti atoms from the measured Gaussian width. As expected, the gas temperatures measured from the Ti spectroscopy are higher than the cell wall temperatures, due to heating by ablation \cite{skoff2011diffusion}.
Typically the gas temperature is 0.4~K to 1.5~K higher, depending on the ablation energy and time after ablation.
We note that, under the conditions explored in this work, the TiHe linewidth is the same as the Ti linewidth to within our experimental error.

\begin{figure}[t]
    \begin{center}
      \includegraphics[width=1.1\linewidth, trim = 55 60 0 50 ]{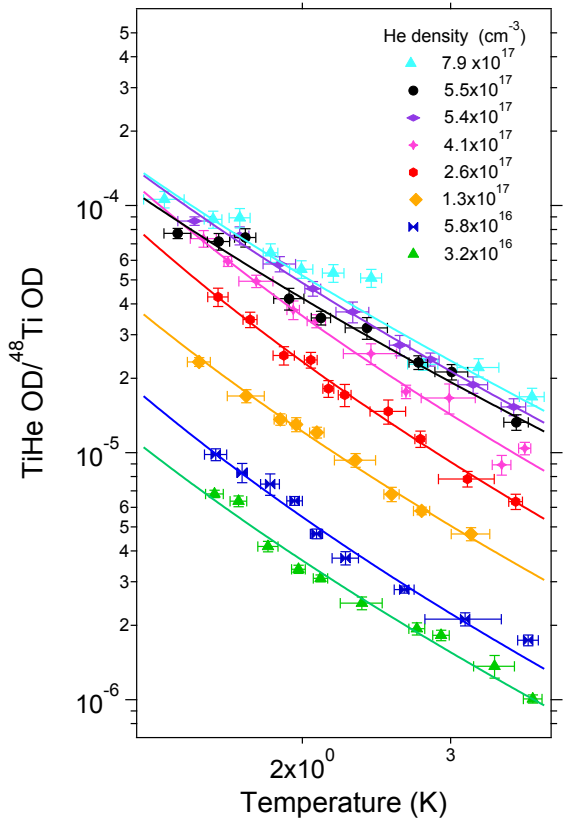} 
    \caption{ \label{fig:TiHeOD_vs_T_all_third}  
        TiHe OD/$^{48}$Ti OD as a function of temperature at 25228.03 cm$^{-1}$. The data was taken at different He densities. $E$ and $c$ are obtained from the fit. 
                  }       
    \end{center}
\end{figure}


Figure~\ref{fig:TiHeOD_vs_TiOD_oneT} shows the dependence of the TiHe density on the Ti density for fixed helium density and ``fixed'' temperature (by selecting a subset of data within a narrow temperature range). 
The expected linear dependence is seen. Importantly, this is independent of the time after ablation, up to the earliest times measured (roughly 0.2~s after ablation, limited by  our detection system). This indicates that the TiHe population reaches thermal equilibrium with the Ti and He atoms on a timescale faster than we are observing, and a timescale faster than the timescale on which temperatures and titanium densities are changing in the cell (due to diffusion and cooling).
Because the TiHe density is proportional to the Ti density at all times we observe, in subsequent analysis we  
consider the ratio TiHeOD/TiOD.

To study the temperature dependence, the TiHeOD/TiOD ratio is measured at fixed helium density and fit to the function 
\begin{equation}\label{f_T}
f(T)= 
cT^{-3/2}e^{-(E/T)}
\end{equation}
as per Eq.~(\ref{eq:equil}). 
%
%
This was repeated at multiple helium densities ranging from $3\times 10 ^{16}$~cm$^{-3}$ to $8\times 10 ^{17}$~cm$^{-3}$. 
Typical data and fits are shown in 
Fig. \ref{fig:TiHeOD_vs_T_all_third}.

\begin{figure}[t]
    \begin{center}
      \includegraphics[width=1.05\linewidth, trim = 85 60 0 30]{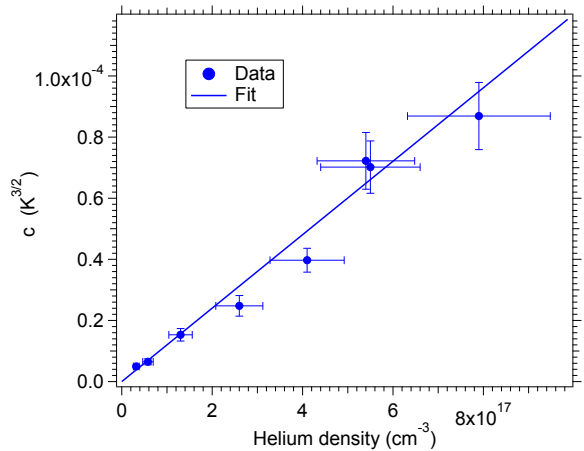} 
    \caption{ \label{fig:c_25228_03} 
           $c$ coefficient vs helium density  for the $^{48}$Ti$^{4}$He transition at $25228.03~$cm$^{-1}$. The fit is to a line $y=bx$.
           }       
    \end{center}
\end{figure}

The energies from these fits show no dependence on the helium density, however the $c$ coefficient of Eq.~(\ref{f_T}) shows a linear dependence on the helium density, as shown in Fig.~\ref{fig:c_25228_03}. This is the expected behavior from Eq. (\ref{eq:equil}), and 
indicates that 
the transitions originate from diatomic molecules, and not trimers or helium clusters.

From a weighted average of the fit values of $E$, we determine the binding energies of the ground-state energy levels corresponding to each transition.
The TiHe transitions that were studied in this experiment and their measured binding energies ($E$) are shown in Table \ref{tab:energyLevels}. These four transitions were the only ones that could be observed over a wide temperature and density range; the signal-to-noise of the remaining molecular transitions made them difficult to measure at high temperatures. Some of these  transitions  could not be identified as due to Ti isotopes and we suspect that they originate  from the $N=1$ ground state, but this could not be determined conclusively due to a lack of knowledge of the excited state structure \cite{TiHe2016sup}. 

The experimental values presented in Table~\ref{tab:energyLevels} have a statistical error of $\pm 0.1$~K. Systematic errors---dominated by the uncertainty in the helium density---add an additional systematic error of $\pm0.25$~K \cite{NancyThesis}. Because the different transitions are measured under the same conditions, the uncertainty in the binding energy differences should be the statistical error of  $\pm0.1$~K.

To theoretically simulate the energy level spectrum of  TiHe, we carry out multichannel bound-state calculations  based on the  Hamiltonian (in atomic units)
\begin{equation}\label{H}
\hat{H} = \frac{-1}{2\mu R}\frac{\partial^2}{\partial R^2}R + \frac{\hat{\bm{N}}^2}{2\mu R^2}+ \hat{H}_\text{SO} + \sum_{\Lambda,\Sigma} V_{\Lambda\Sigma}(R) |\Lambda\Sigma\rangle \langle \Lambda\Sigma|,
\end{equation}
where $R$ is the Ti-He internuclear distance, $\mu$ is the reduced mass \cite{TiHe2016sup}, $\hat{\bm{N}}$ is the rotational angular momentum of the complex,  $\hat{H}_\text{SO}=A\hat{\bm{L}}_a\cdot\hat{\bm{S}}_a$ is the spin-orbit (SO) interaction, which couples the electronic orbital and spin angular momenta of Ti to form the resultant $\hat{\bm{J}}_a$, and $A$ is the SO constant. The Ti-He interaction potential is expanded in terms of the adiabatic electronic basis functions $|\Lambda\Sigma\rangle$ where $\Lambda$ and $\Sigma$ are the projections of $\hat{\bm{L}}_a$ and $\hat{\bm{S}}_a$ on the molecular axis. 

To obtain the most accurate parametrization of the Hamiltonian (\ref{H}), we carried out new high-level {\it ab initio} 
calculations of the adiabatic potentials $V_{\Lambda\Sigma}(R)$ using the state-averaged complete active space self-consistent field (CASSCF) method \cite{klos04TiHe} to obtain reference states with proper $\Lambda$, followed by internally-contracted multireference self-consistent field calculations ~\cite{werner:1988} with single and double excitations and a Davidson correction~\cite{langhoff:1974} ({\em ic}-MRCISD+Q) to account for higher excitations.

\begin{table}
\caption{The calculated energy levels of $^{48}$Ti$^4$He (in K) and the corresponding Hund's case (e) quantum numbers ($J_a=2$ for all levels), along with the experimentally measured binding energies  of the four largest observed transitions. The  experimental statistical error bars are $\pm 0.1$~K, with an additional common systematic error of $\pm 0.25$ K. 
\label{tab:energyLevels}}

\begin{center}

\begin{tabular}{ccc c||cc}
\multicolumn{3}{c}{Theory}  &  & \multicolumn{2}{c}{Experiment} \\
\hline
$N$ &  $J$ & $E$  (K) & & Frequency (cm$^{-1}$)  & $E$ (K) \\
\hline
0 & $ 2 $ &  $-1.8244$  & & 25228.15&  -1.95 \\
 1 & $ 1 $ & $-1.5292$  & & 25228.10& -1.33 \\
 1 & $ 3 $ & $-1.5079$  & & 25228.03&  -1.52 \\
 1 & $ 2 $  &  $-1.4689$ & & 25227.99& -1.40 \\
\end{tabular}

\end{center}
\end{table}

For an improved description of the Ti-He interaction energy, we use  a quintuple-zeta basis set (aug-cc-pwcv5Z-DK) specifically designed for Douglass-Kroll integrals in all-electron scalar relativistic calculations, augmented additionally with a  $3s3p2d2f1g1h$ set of mid-bond functions~\cite{klos04TiHe}. This basis is much larger than used in the previous {\it ab initio} calculations \cite{klos04TiHe,krems05TiHe}. We correct the interaction energies for the basis set superposition error and for  size-consistency at $R=500 a_0$.
The new potentials are uniformly more attractive by $\sim$1.3 cm$^{-1}$ than the previous potentials due to a larger basis set used in this work and the inclusion of the mid-bond functions. The new potential minima occur at shorter values of $R$ (by $\sim$0.2-0.3~$a_0$); however, the energy order of the new $V_{\Lambda\Sigma}$ potentials is the same as obtained previously  \cite{klos04TiHe,krems05TiHe}. 

We obtain the bound-state energy levels  of TiHe by diagonalizing the Hamiltonian (\ref{H}) expressed in the basis of direct products of Hund's case (e) basis functions \cite{TiHe2016sup}  obtained by vector coupling of $\hat{\bm{J}_a}$ and $\hat{\bm{N}}$  to form the total angular momentum of the complex $\hat{\bm{J}}$ and the radial basis functions in the discrete variable representation \cite{ColbertMiller1992}. The calculated bound-state energies are converged to $< 1$\%.

\begin{figure}[t]
    \begin{center}
      \includegraphics[width=\linewidth, trim = 0 112 0 100]{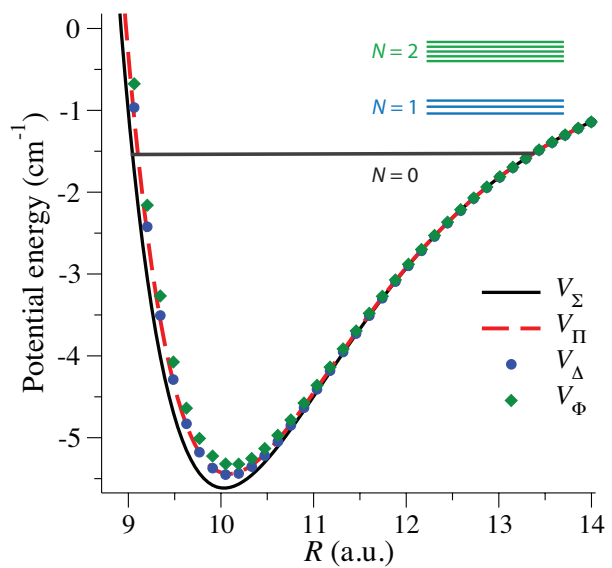}
    \caption{ \label{fig:potentials} 
         {\it Ab initio} potentials and bound levels of the TiHe vdW complex. The anisotropy-induced splitting of the rotational levels is exaggerated for clarity.   }       
    \end{center}
\end{figure}

Figure \ref{fig:potentials} shows our {\it ab initio} interaction potentials for the TiHe electronic states of $\Sigma,\Pi, \Delta$, and $\Phi$ symmetries ($\Lambda=0{-}3$). The potentials  are nearly degenerate at all $R$ due to the suppressed electronic anisotropy of the ground-state Ti electronic configuration ($3d^24s^2$) \cite{klos04TiHe,krems05TiHe}.
Analysis of molecular eigenvectors shows a very small amount of mixing between the different $N$-states, which is expected since the anisotropic terms are small compared to the splitting between the rotational levels.  Thus, $N$ is a good quantum number for the lowest bound states of Ti$(^3F)$-He, and its rotational energy level structure (see Fig.~\ref{fig:potentials}) can be fully understood in terms of the rigorously conserved quantum numbers  $J_a$, $J$, and $N$, a nearly perfect example of Hund's case (e) coupling scheme.

  The calculated energies of the lowest rotational states of TiHe are listed in Table I. The ground rotational energy level ($N=0$) is non-degenerate and has a binding energy of 1.82~K.  The excited rotational states  in different $N$ manifolds are split by the interaction anisotropy into $(2N+1)$ sublevels with $J=|J_a-N|,\ldots,J_a+N$.
The calculated binding energies are in excellent agreement with experiment for $N=0$ and $N=1$ (see Table~\ref{tab:energyLevels}), demonstrating that our {\it ab initio} potentials  are highly accurate.  The levels in the $N=2$ and $N=3$ manifolds  were  not observed experimentally.  This is expected: the  selection rules for the electronic transitions between Hund's case (e) levels allow only $q$-branch transitions \cite{TiHe2016sup}. The excited state rotational manifold is less weakly bound by $\sim 1$~cm$^{-1}$ (determined from the blueshift) so it is unlikely to support $N=2$ or $N=3$ rotational levels.



In summary, we have observed the formation of cold TiHe vdW molecules featuring an exotic angular momentum coupling scheme---Hund's case (e)---arising from the anisotropic nature of the ground-state Ti-He interaction. The molecules were detected in their ground and first excited  rotational states and our thermodynamic measurements of their binding energies are  in quantitative agreement with theoretical calculations based on highly accurate {\it ab initio} interaction potentials. Our results show that the ground and rotationally excited TiHe molecules can form in cryogenic He buffer gas,  opening up the possibility of studying three-body recombination and non-universal physics   in the multiple partial wave regime  \cite{WangEsryNJP2011,PhysRevA.80.062702}.  The chemical reactions,   inelastic scattering, and Zeeman predissociation of cold vdW molecules can now be investigated experimentally and possibly controlled with external electromagnetic fields \cite{Krems04}.
This work could also be extended to explore the formation of larger clusters (such as TiHe$_2$ trimers) at higher helium densities.

%
%
This material is based upon work supported by the  National Science Foundation under Grant No. PHY 1265905. J.K. acknowledges support from the NSF Grant No. CHE-1565872 to Millard Alexander. 




%

\end{document}